\documentclass[12pt,dvips]{article}

\usepackage{cite,epsfig}
\usepackage{color}
\definecolor{GreenYellow}   {cmyk}{0.15,0,0.69,0}
\definecolor{Yellow}        {cmyk}{0,0,1,0}
\definecolor{Goldenrod}     {cmyk}{0,0.10,0.84,0}
\definecolor{Dandelion}     {cmyk}{0,0.29,0.84,0}
\definecolor{Apricot}       {cmyk}{0,0.32,0.52,0}
\definecolor{Peach}         {cmyk}{0,0.50,0.70,0}
\definecolor{Melon}         {cmyk}{0,0.46,0.50,0}
\definecolor{YellowOrange}  {cmyk}{0,0.42,1,0}
\definecolor{Orange}        {cmyk}{0,0.61,0.87,0}
\definecolor{BurntOrange}   {cmyk}{0,0.51,1,0}
\definecolor{Bittersweet}   {cmyk}{0,0.75,1,0.24}
\definecolor{RedOrange}     {cmyk}{0,0.77,0.87,0}
\definecolor{Mahogany}      {cmyk}{0,0.85,0.87,0.35}
\definecolor{Maroon}        {cmyk}{0,0.87,0.68,0.32}
\definecolor{BrickRed}      {cmyk}{0,0.89,0.94,0.28}
\definecolor{Red}           {cmyk}{0,1,1,0}
\definecolor{OrangeRed}     {cmyk}{0,1,0.50,0}
\definecolor{RubineRed}     {cmyk}{0,1,0.13,0}
\definecolor{WildStrawberry}{cmyk}{0,0.96,0.39,0}
\definecolor{Salmon}        {cmyk}{0,0.53,0.38,0}
\definecolor{CarnationPink} {cmyk}{0,0.63,0,0}
\definecolor{Magenta}       {cmyk}{0,1,0,0}
\definecolor{VioletRed}     {cmyk}{0,0.81,0,0}
\definecolor{Rhodamine}     {cmyk}{0,0.82,0,0}
\definecolor{Mulberry}      {cmyk}{0.34,0.90,0,0.02}
\definecolor{RedViolet}     {cmyk}{0.07,0.90,0,0.34}
\definecolor{Fuchsia}       {cmyk}{0.47,0.91,0,0.08}
\definecolor{Lavender}      {cmyk}{0,0.48,0,0}
\definecolor{Thistle}       {cmyk}{0.12,0.59,0,0}
\definecolor{Orchid}        {cmyk}{0.32,0.64,0,0}
\definecolor{DarkOrchid}    {cmyk}{0.40,0.80,0.20,0}
\definecolor{Purple}        {cmyk}{0.45,0.86,0,0}
\definecolor{Plum}          {cmyk}{0.50,1,0,0}
\definecolor{Violet}        {cmyk}{0.79,0.88,0,0}
\definecolor{RoyalPurple}   {cmyk}{0.75,0.90,0,0}
\definecolor{BlueViolet}    {cmyk}{0.86,0.91,0,0.04}
\definecolor{Periwinkle}    {cmyk}{0.57,0.55,0,0}
\definecolor{CadetBlue}     {cmyk}{0.62,0.57,0.23,0}
\definecolor{CornflowerBlue}{cmyk}{0.65,0.13,0,0}
\definecolor{MidnightBlue}  {cmyk}{0.98,0.13,0,0.43}
\definecolor{NavyBlue}      {cmyk}{0.94,0.54,0,0}
\definecolor{RoyalBlue}     {cmyk}{1,0.50,0,0}
\definecolor{Blue}          {cmyk}{1,1,0,0}
\definecolor{Cerulean}      {cmyk}{0.94,0.11,0,0}
\definecolor{Cyan}          {cmyk}{1,0,0,0}
\definecolor{ProcessBlue}   {cmyk}{0.96,0,0,0}
\definecolor{SkyBlue}       {cmyk}{0.62,0,0.12,0}
\definecolor{Turquoise}     {cmyk}{0.85,0,0.20,0}
\definecolor{TealBlue}      {cmyk}{0.86,0,0.34,0.02}
\definecolor{Aquamarine}    {cmyk}{0.82,0,0.30,0}
\definecolor{BlueGreen}     {cmyk}{0.85,0,0.33,0}
\definecolor{Emerald}       {cmyk}{1,0,0.50,0}
\definecolor{JungleGreen}   {cmyk}{0.99,0,0.52,0}
\definecolor{SeaGreen}      {cmyk}{0.69,0,0.50,0}
\definecolor{Green}         {cmyk}{1,0,1,0}
\definecolor{ForestGreen}   {cmyk}{0.91,0,0.88,0.12}
\definecolor{PineGreen}     {cmyk}{0.92,0,0.59,0.25}
\definecolor{LimeGreen}     {cmyk}{0.50,0,1,0}
\definecolor{YellowGreen}   {cmyk}{0.44,0,0.74,0}
\definecolor{SpringGreen}   {cmyk}{0.26,0,0.76,0}
\definecolor{OliveGreen}    {cmyk}{0.64,0,0.95,0.40}
\definecolor{RawSienna}     {cmyk}{0,0.72,1,0.45}
\definecolor{Sepia}         {cmyk}{0,0.83,1,0.70}
\definecolor{Brown}         {cmyk}{0,0.81,1,0.60}
\definecolor{Tan}           {cmyk}{0.14,0.42,0.56,0}
\definecolor{Gray}          {cmyk}{0,0,0,0.50}
\definecolor{Black}         {cmyk}{0,0,0,1}
\definecolor{White}         {cmyk}{0,0,0,0}

\usepackage{axodraw}

\newcommand{\imag}{\Im {\rm m}}
\newcommand{\real}{\Re {\rm e}}

\def\lsim{\:\raisebox{-0.5ex}{$\stackrel{\textstyle<}{\sim}$}\:}
\def\gsim{\:\raisebox{-0.5ex}{$\stackrel{\textstyle>}{\sim}$}\:}

\begin{document}

\begin{flushright}
CERN-PH-TH/2006-102 \\
MAN/HEP/2006/18\\
hep-ph/0605288 \\
May 2006
\end{flushright}

\begin{center}
{\bf {\LARGE Higgs Phenomenology with 
                {\color{Red}CP}{\color{Blue}super}{\color{OliveGreen}H} }}
\end{center}


\begin{center}
{\large John Ellis$^{\,a}$, Jae Sik Lee$^{\,b}$
and Apostolos Pilaftsis$^{a,c}$}
\end{center}

\begin{center}
{\em $^a$Theory Division, Physics Department, CERN, CH-1211 Geneva 23,
Switzerland}\\[2mm]
{\em $^b$Center for Theoretical Physics, School of Physics, Seoul National University,}\\
{\em Seoul 151-747, Korea}\\[2mm]
{\em $^c$School of Physics and Astronomy, University of Manchester}\\
{\em Manchester M13 9PL, United Kingdom}
\end{center}

\bigskip\bigskip\bigskip


\centerline{\bf ABSTRACT}
\noindent
The  MSSM  contains  CP-violating   phases  that  may  have  important
observable effects  in Higgs physics.  We review  recent highlights in
Higgs phenomenology  obtained with the  code {\tt CPsuperH},  a useful
tool  for studies of  the production,  mixing and  decay of  a coupled
system of  the neutral  Higgs bosons at  future high  energy colliders
such as the LHC, ILC ($\gamma$LC), and a muon collider (MC). {\tt CPsuperH}
implements  the  constraints  from  upper limits  on  electric  dipole
moments, and  may be extended to include  other related low-energy observables,  
such   as  $b\rightarrow  s\gamma$   and  $B
\rightarrow  K  l\,l$, and  to  compute  the  relic abundance  of  the
lightest neutralino.

\newpage


\section{Introduction}

CP violation is one of the most sensitive probes for new physics
beyond the Standard Model, and some additional source of CP violation
is needed to account for cosmological baryogenesis.  Ongoing
experiments provide no indications of any flavour or CP-violating
physics beyond the CKM mixing paradigm.  Nevertheless, $B$-factories
and the Tevatron collider are constantly refining their probes of the
CKM mechanism, and the LHC and other TeV-scale colliders will open up
new vistas in the study of CP violation.  Supersymmetry is one of the
more attractive extensions of the Standard Model, and there are many
possible supplementary sources of CP violation even within the minimal
supersymmetric extension of the Standard Model (MSSM). These are
reduced to just two if one assumes universal soft
supersymmetry-breaking parameters, which would be sufficient to
facilitate baryogenesis at the electroweak scale. They would also have
a wealth of implications for many different areas in particle-physics
phenomenology, notably through radiative corrections in the Higgs
sector ~\cite{Pilaftsis:1998pe,Pilaftsis:1998dd,Pilaftsis:1999qt,
Demir,Choi:2000wz,Carena:2000yi,Carena:2001fw}.  
CP-violating effects may play a prominent role in Higgs production at future 
high energy colliders, such as the
LHC~\cite{Bernreuther:1993df,Bernreuther:1993hq,Dedes1,Dedes2,CL_LHC,CKL_LHC,
Carena:2002bb,Arhrib,Christova1,Christova2,Khater,Borzumati:2004rd,CFLMP,Akeroyd,
Ghosh:2004cc,Ghosh:2004wr,KMR,Borzumati:2006zx}, 
ILC~\cite{Grzadkowski,AkeroydArhrib}, 
$\gamma$LC~\cite{CL_PLC,Lee:2001ru,CCKL,Choi:2004ne,Heinemeyer,ACHL,Godbole,
Asakawa,Choi:2004kq}
and MC~\cite{Atwood,Grzadkowski_mumu,Pilaftsis_mumu,CL_mumu,ACL_mumu,CDGL,
Berger,Blochinger:2002hj,Bernabeu:2006zs}.  
In addition, CP-violating phenomena mediated by Higgs-boson exchanges may
manifest themselves in a number of low-energy observables such as
electric dipole moments
~\cite{Ibrahim:1998je,Ibrahim:1999af,Brhlik:1999ub,Brhlik:1999pw,
Pokorski:1999hz,Accomando:1999uj,Bartl:1999bc,Falk:1999tm,
Abel:2001vy,Demir:2002gg,Feng:2003mg,Bartl:2003ju,
Chang:1998uc,Pilaftsis:2002fe,Ibrahim:2003jm,Semertzidis:1999kv}.
The CP phases of the MSSM may also affect flavour-changing
neutral-current processes and CP asymmetries involving $K$ and $B$
mesons
~\cite{Farley:2003wt,Huang:1998bf,Choudhury:1998ze,Huang:1998vb,
Hamzaoui:1998nu,Babu:1999hn,
Bobeth:2001sq,Isidori:2001fv,Dedes:2001fv,Arnowitt:2002cq,Baek:2002wm,
Buras:2001mb,Buras:2002vd,D'Ambrosio:2002ex,Mizukoshi:2002gs,
Chankowski:2000ng,Huang:2000sm,Demir:2001yz,Boz:2002wa,
Ibrahim:2002fx,Ibrahim:2003ca,Demir:2003bv,Carena:1999py,Dedes:2002er}.
Moreover, CP-violating Higgs effects may influence the annihilation
rates of cosmic relics and hence the abundance of dark matter in the
Universe ~\cite{Gondolo:1999gu,Falk:1999mq,Choi:2000kh,Nihei:2005va}.
Finally, we recall that an accurate determination of the Higgs
spectrum in the presence of CP violation is crucial for testing the
viability of electroweak baryogenesis in the MSSM
~\cite{Carena:1997ki,Laine:1998qk,Funakubo:1999ws,Grant:1998ci,Losada:1999tf,
Cline:2000nw,Carena:2000id,Laine:2000rm,Kainulainen:2001cn,Balazs:2004ae,
Konstandin:2005cd,Cirigliano:2006wh,Cirigliano:2006dg}

The {\tt Fortran} code {\tt CPsuperH} \cite{Lee:2003nt} is a
powerful and efficient computational tool for understanding
quantitatively such phenomenological subjects
within the framework of the MSSM with explicit CP
violation. 
It calculates  the mass  spectrum
and  decay  widths of  the neutral  and
charged Higgs bosons in the most general MSSM including CP-violating phases.   
In addition, it computes all the couplings of the 
neutral Higgs bosons $H_{1,2,3}$ and the charged Higgs boson $H^+$. 
The    program   is   based   on   
the results obtained in Refs.~\cite{Choi:1999uk,Choi:2001pg,Choi:2002zp}
and the most recent
renormalization-group-improved effective-potential approach, which includes
dominant   higher-order   logarithmic   and  threshold   corrections,
$b$-quark   Yukawa-coupling   resummation   effects,  and   
Higgs-boson pole-mass shifts \cite{Carena:2000yi,Carena:2001fw}
The masses and couplings  of the charged and neutral Higgs bosons
are computed at a similar  high-precision level.
Even in the CP-conserving case, {\tt CPsuperH} is unique in
computing the neutral and charged Higgs-boson couplings and masses
with equally high levels of precision, and is therefore a useful
tool for the study of MSSM Higgs phenomenology at present and future
colliders.

\section{{\color{Red}CP}{\color{Blue}super}{\color{OliveGreen}H}}
 
The tarred and gzipped program file {\tt CPsuperH.tgz} can be
downloaded from
\footnote{Some new features appearing in this write-up, for example,
the propagator matrix {\tt DH(3,3)} and some low-energy observables,
will be implemented in a forthcoming version of {\tt CPsuperH}.}:
\begin{center}
{\tt http://www.hep.man.ac.uk/u/jslee/CPsuperH.html}
\end{center}
When running {\tt CPsuperH}, the main numerical output is stored in
arrays.  These include the masses of the three neutral Higgs bosons,
labelled in order of increasing mass such that $M_{H_1}\le M_{H_2} \le
M_{H_3}$.  Since the neutral pseudoscalar Higgs bosons mixes with the
neutral scalars in the presence of CP violation, the charged 
Higgs-boson pole mass $M_{H^\pm}$ is used as an input parameter.  {\tt CPsuperH}
also yields the $3\times 3$ Higgs mixing matrix, $O_{\alpha i}$:
$(\phi_1,\phi_2,a)^T_\alpha=O_{\alpha i}(H_1,H_2,H_3)^T_i$ and all the
couplings of the neutral and charged Higgs bosons. These include Higgs
couplings to leptons, quarks, neutralinos, charginos, stops, sbottoms,
staus, tau sneutrinos, gluons, photons, and massive vector bosons, as
well as Higgs-boson self-couplings.  We note that, in addition to
quantities in the Higgs sector, including decay widths and branching
fractions, {\tt CPsuperH} also calculates and stores the masses and
mixing matrices of the stops, sbottoms, staus, charginos, and
neutralinos. For a full description, we refer to
Ref.~\cite{Lee:2003nt}.
 
\section{Collider Signatures}

To analyze CP-violating phenomena  in the production, mixing and decay
of a  coupled system~\cite{Pilaftsis:1997dr} of  multiple CP-violating
MSSM neutral Higgs bosons at colliders, we need a ``full" 3 $\times$ 3
propagator matrix $D(s)$, given by \cite{Ellis:2004fs}
\begin{equation}
  \label{eq:Hprop}
D (\hat{s}) = \hat{s}\,
\left(\begin{array}{ccc}
       \hat{s}-M_{H_1}^2+i\Im{\rm m}\widehat{\Pi}_{11}(\hat{s}) &
i\Im{\rm m}\widehat{\Pi}_{12}(\hat{s})&
       i\Im{\rm m}\widehat{\Pi}_{13}(\hat{s}) \\
       i\Im{\rm m}\widehat{\Pi}_{21}(\hat{s}) &
\hat{s}-M_{H_2}^2+i\Im{\rm m}\widehat{\Pi}_{22}(\hat{s})&
       i\Im{\rm m}\widehat{\Pi}_{23}(\hat{s}) \\
       i\Im{\rm m}\widehat{\Pi}_{31}(\hat{s}) & i\Im{\rm m}\widehat{\Pi}_{32}(\hat{s}) &
       \hat{s}-M_{H_3}^2+
       i\Im{\rm m}\widehat{\Pi}_{33}(\hat{s})
      \end{array}\right)^{-1} \,,
\end{equation}
where $\hat{s}$ is  the center-of-mass energy squared, $M_{H_{1,2,3}}$
are the one-loop Higgs-boson pole  masses, and the absorptive parts of
the   Higgs   self-energies  $\Im{\rm   m}\widehat{\Pi}_{ij}(\hat{s})$
receive   contributions  from  loops   of  fermions,   vector  bosons,
associated pairs  of Higgs and  vector bosons, Higgs-boson  pairs, and
sfermions.  The computed  resummed  Higgs-boson propagator
matrix has been stored in the array {\tt DH(3,3)}.

The so called tri-mixing scenario has been taken for studying the
production, mixing and decay of a coupled system of the neutral Higgs
bosons at colliders. This scenario is characterized by a large value
of $\tan\beta=50$ and the light charged Higgs boson
$M_{H^\pm}=155$ GeV. All the three-Higgs states mix
significantly in this scenario in the presence of CP-violating
mixing. Without CP violation, only two CP-even states mix.  For
details of the scenario, see Refs.~
\cite{Ellis:2004fs,Ellis:2004hw,Ellis:2005fp,Ellis:2005ik}.

\subsection{LHC}

At the LHC, the matrix element for the process 
$g(\lambda_1)g(\lambda_2)\rightarrow H \rightarrow f(\sigma)\bar{f}(\bar{\sigma})$ can
conveniently be represented by the helicity amplitude
\begin{equation}
{\cal M}^{gg}(\sigma\bar{\sigma};\lambda_1\lambda_2)\ =\
\frac{\alpha_s\, g_f\,\sqrt{\hat{s}}\,\delta^{ab}}{4\pi v }
\langle\sigma;\lambda_1\rangle_g
\delta_{\sigma\bar{\sigma}}\delta_{\lambda_1\lambda_2}\,,
\label{eq:ggamp}
\end{equation}
where $a$ and $b$ are indices of the  SU(3) generators in the  
adjoint representation and $\sigma$, $\bar{\sigma}$, and $\lambda_{1,2}$
denote the helicities of fermion, antifermion, and gluons, respectively.
The amplitude $\langle\sigma;\lambda\rangle_g$ is defined as
\begin{equation}
\langle\sigma;\lambda\rangle_g \equiv \sum_{i,j=1,2,3}
[S_i^g(\sqrt{\hat{s}})+i\lambda P_i^g(\sqrt{\hat{s}})]\:
D_{ij}(\hat{s})\: (\sigma\beta_f g_{H_j\bar{f}f}^S -i
g_{H_j\bar{f}f}^P)\,,
\end{equation}
where
\begin{eqnarray}
S^g_i(\sqrt{\hat{s}})&=&\sum_{f=b,t}
g_fg^{S}_{H_i\bar{f}f}\,\frac{v}{m_f} F_{sf}(\tau_{f})
-\sum_{\tilde{f}_j=\tilde{t}_1,\tilde{t}_2,\tilde{b}_1,\tilde{b}_2}
g_{H_i\tilde{f}^*_j\tilde{f}_j}
\frac{v^2}{4m_{\tilde{f}_j}^2} F_0(\tau_{\tilde{f}_j}) \,, \nonumber \\
P^g_i(\sqrt{\hat{s}})&=&\sum_{f=b,t}
g_fg^{P}_{H_i\bar{f}f}\,\frac{v}{m_f}
F_{pf}(\tau_{f}) \,,
\end{eqnarray}
where $\beta_f=\sqrt{1-4m_f^2/\hat{s}}$,
$\tau_{x}=\hat{s}/4m_x^2$ and  $\hat{s}$-dependent  $S^g_i$ and
$P^g_i$  are  scalar   and  pseudoscalar  form  factors\footnote{These
$\hat{s}$-dependent gluon-gluon-Higgs  couplings are stored  in arrays
{\tt SGLUE(3)} and {\tt PGLUE(3)}.}, respectively.
The Higgs-boson couplings to quarks $g^{S\,,P}_{H_i\bar{f}f}$
and squarks $g_{H_i\tilde{f}^*_j\tilde{f}_j}$, and the explicit forms of
the functions $F_{sf,pf,0}$ are coded in {\tt CPsuperH} \cite{Lee:2003nt}.
When $f=\tau\,,t\,,\chi^0\,,\chi^\pm$, etc, one can construct CP asymmetries in
the longitudinal and/or transverse polarizations of the final fermions which
can be observed at the LHC.
The CP asymmetries can be defined similarly in the case of 
other production mechanisms such as $b$-quark and weak-boson fusions.
 
\begin{figure}[t]
\vspace{-1.6cm}
\centerline{\epsfig{figure=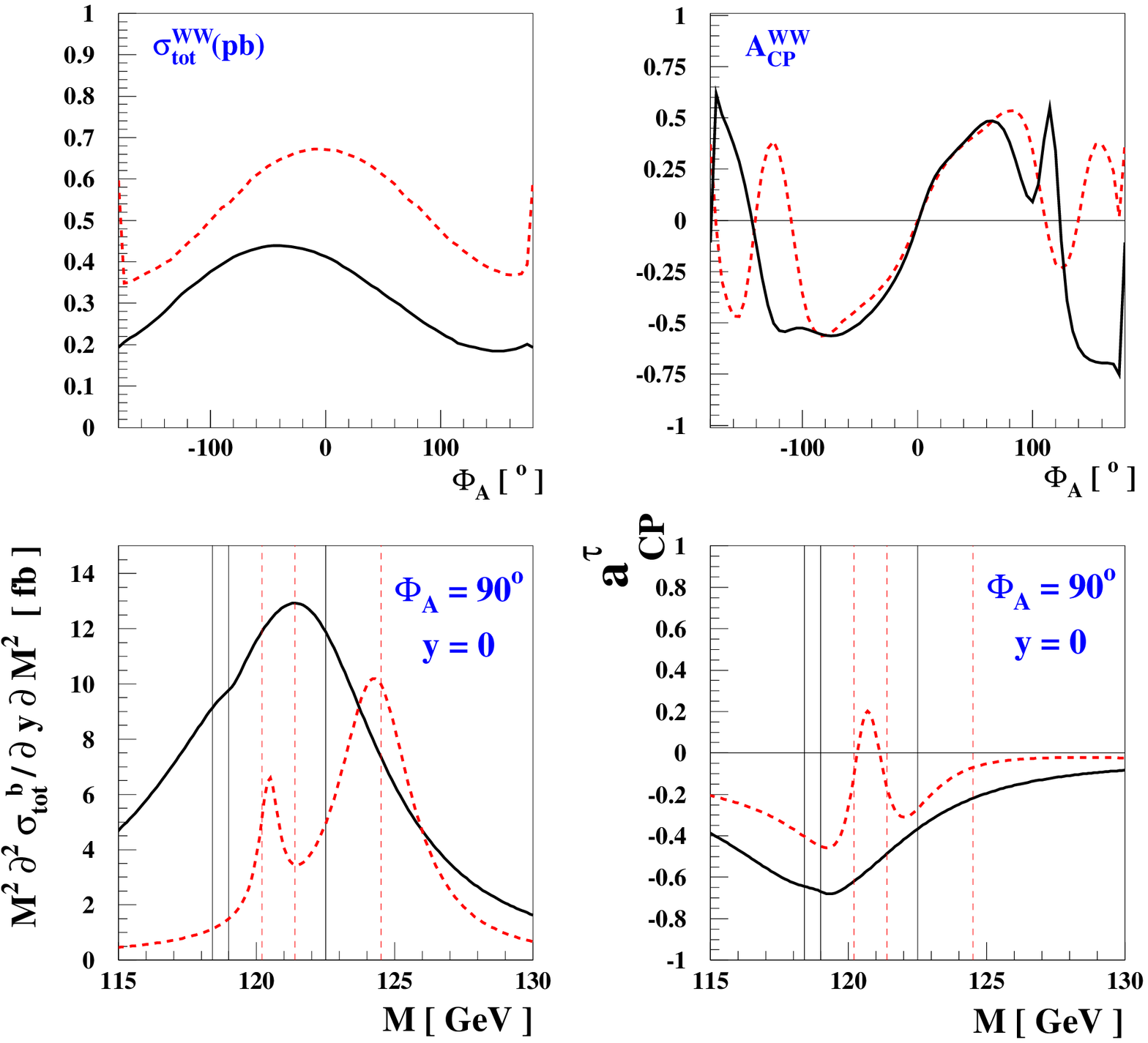,height=13cm,width=16cm}}
\vspace{-0.5cm}
\caption{\small \it The cross section 
$\sigma^{WW}_{\rm tot}[pp(WW) \to \tau^+\tau^- X]$
(upper-left panel) at the LHC and its associated total CP asymmetry 
${\cal A}^{WW}_{\rm CP}
\equiv [\sigma^{WW}_{RR}-\sigma^{WW}_{LL}]/[\sigma^{WW}_{RR}+\sigma^{WW}_{LL}]$ 
(upper-right panel) as functions of 
$\Phi_A=\Phi_{A_t} =\Phi_{A_b} =\Phi_{A_\tau}$,
where $\sigma^{WW}_{RR,LL}\equiv\sigma(pp\,(WW) \to \tau_{R,L}^+ \tau_{R,L}^- X)$. 
We have considered a
tri-mixing scenario with $\Phi_3=-10^\circ$ (dotted lines) and $-90^\circ$ (solid lines). 
For details of the scenario and the CP asymmetry, see \cite{Ellis:2004fs}.
The lower frames are for Higgs bosons produced in diffractive collisions at the LHC.
The lower-left frame shows the hadron-level cross sections when the 
Higgs bosons decay into $b$ quarks, as functions of the
invariant mass $M$ with $\Phi_A=90^\circ$ and rapidity $y=0$ . The vertical lines  indicate the three Higgs-boson  pole-mass positions.
The lower-right frame shows the CP-violating asymmetry when the 
Higgs bosons decay into $\tau$ leptons.
See \cite {Ellis:2005fp} for details.  }
\vspace{-0.5cm}
\label{fig:lhc}
\end{figure}

When $\tan\beta$ is large, $b$-quark fusion is an important mechanism for producing
neutral Higgs bosons.
In the centre-of-mass coordinate system for the $b {\bar b}$ pair, the
helicity amplitudes for the process
$b(\lambda)\,\bar{b}(\bar{\lambda})\rightarrow H \rightarrow f(\sigma)\bar{f}(\bar{\sigma})$ 
are given by
\begin{equation}
{\cal M}^{b\bar{b}}(\sigma\bar{\sigma};\lambda\bar{\lambda})\ =\
-\frac{g\, m_b\, g_f}{2 M_W}\,
\langle \sigma ; \lambda \rangle_{b}
\delta_{\sigma\bar{\sigma}}\delta_{\lambda\bar{\lambda}}\,,
\label{eq:bbamp}
\end{equation}
where
\begin{equation}
\langle \sigma ; \lambda \rangle_{b}\  \equiv\ \sum_{i,j=1,2,3}
(\lambda\beta_b\,g^S_{H_i\bar{b}b}+i g^P_{H_i\bar{b}b})\: D_{ij}(\hat{s})\:
(\sigma\beta_f\,g^S_{H_j\bar{f}f}-i g^P_{H_j\bar{f}f})\,.
\end{equation}
For the tri-mixing scenario, $b$-quark fusion is the main production mechanism,
and dilutes the possibly large CP asymmetry in longitudinally-polarized
gluon fusion when $f=\tau$.  This dilution is basically due to the dominance of the
$b$-quark loop contribution to the self energies and the experimental impossibility of
distinguishing gluon and $b$-quark fusion events.
In this case, the most promising channel for probing Higgs-sector CP violation
may be the weak-boson fusion process, which can be separated from $gg$ and
$b\bar{b}$ collisions by applying a number
of kinematic  cuts~\cite{Asai:2004ws}, including the imposition of  a veto on
any  hadronic  activity  between the jets~\cite{Cahn,BargerHan,Iordanidis,Rainwater,JF1}:
$W^+W^- \to  H_{1,2,3} \to \tau^+ \tau^-$. The cross section
$\sigma[pp(W^+W^-) \to H \to \tau^+ \tau^- X]$ lies between 0.2 and 0.6 pb and 
the CP asymmetry is large for a wide
range of CP phases, see the two upper frames of Fig.~\ref{fig:lhc}.

Higgs-boson   production in an   exclusive  diffractive collision $p+p
\rightarrow p+H_i+p$
offers  unique  
possibilities  for exploring Higgs    physics in ways  that  would  be
difficult or even  impossible   in  inclusive  Higgs  production \cite {Ellis:2005fp}.
In spite of the low and theoretically uncertain luminosity of the process,
what makes diffraction
so attractive compared to the inclusive processes are the clean environment due to the
large rapidity gap and the good Higgs-mass resolution  of  the
order of 1 GeV which may  be achievable by precise measurements
of the momenta  of  the  outgoing  protons in   detectors a long   way
downstream from  the interaction point.  It may be possible to disentangle
nearly-degenerate Higgs bosons  by examining the  production lineshape
of the coupled system of neutral Higgs bosons, see lower-left frame of Fig.~\ref{fig:lhc}.  
Moreover, the CP-odd polarization asymmetry can be measured when 
the polarization information of Higgs decay products is available,
see the lower-right frame of Fig.~\ref{fig:lhc}.

\subsection{ILC}
\begin{figure}[tbh]
\hspace{1.0cm}
\begin{picture}(500,100)(0,0)
\SetWidth{1.3}
\ArrowLine(0,100)(40,50)\Text(-20,100)[]{$e^-$}
\ArrowLine(40,50)(0,0)\Text(-20,10)[]{$e^+$}
\Photon(40,50)(80,50){4}{6}\Text(60,70)[]{$Z^*$}
\Photon(80,50)(120,100){4}{5}\Text(130,95)[]{$Z$}
\DashLine(80,50)(120,0){3}\Text(130,5)[]{$H_i$}
\Text(60,-15)[]{(a)}
\ArrowLine(170,100)(210,80)
\ArrowLine(210,20)(170,0)
\ArrowLine(210,80)(250,100)\Text(265,100)[]{$\nu_e$}
\ArrowLine(250,0)(210,20)\Text(265,0)[]{$\overline{\nu_e}$}
\Photon(210,80)(230,50){4}{6}\Text(240,70)[]{$W^-$}
\Photon(210,20)(230,50){4}{6}\Text(240,30)[]{$W^+$}
\DashLine(230,50)(260,50){3}\Text(270,50)[]{$H_i$}
\Text(220,-15)[]{(b)}
\ArrowLine(300,100)(340,50)
\ArrowLine(340,50)(300,0)
\Photon(340,50)(380,50){4}{6}\Text(360,70)[]{$Z^*$}
\DashLine(380,50)(410,0){3}\Text(420,95)[]{$H_i$}
\DashLine(380,50)(410,100){3}\Text(420,5)[]{$H_j$}
\Text(360,-15)[]{(c)}
\end{picture}
\vspace{0.0 cm}
\caption{\small \it Three main production mechanisms of the neutral Higgs bosons at the 
ILC: (a) Higgsstrahlung, (b) $WW$ fusion, and (c) pair production.}
\vspace{-0.5 cm}
\label{fig:higgs_ilc}
\end{figure}
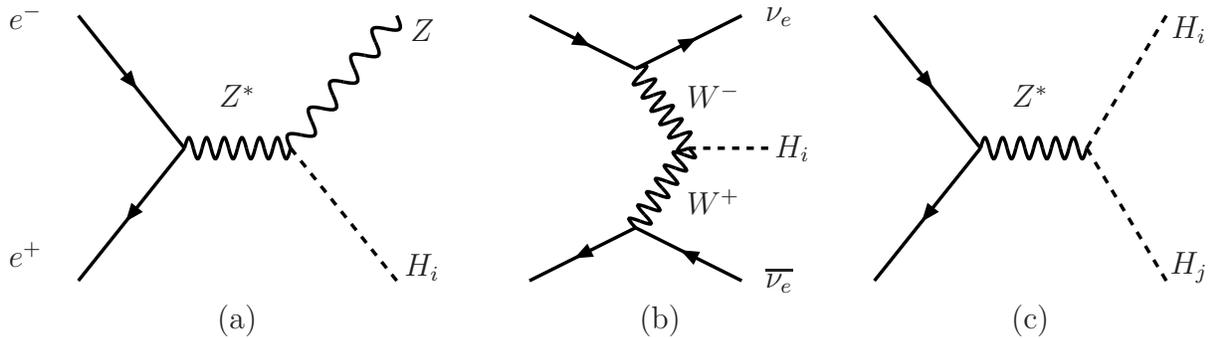

A future $e^+e^-$ linear collider, such as the projected ILC, will
have the potential to probe the Higgs sector with higher precision
than the LHC.  At the ILC, the Higgs-boson coupling to a pair of
vector bosons, $g_{H_iVV}=c_\beta O_{\phi_1 i}+s_\beta O_{\phi_2 i}$,
plays a crucial role. There are three main processes for producing the
neutral Higgs bosons: Higgsstrahlung, $WW$ fusion, and pair
production, see Fig.~\ref{fig:higgs_ilc}.

\vspace{0.3 cm}

\begin{figure}[t]
\vspace{-1.3cm}
\centerline{\epsfig{figure=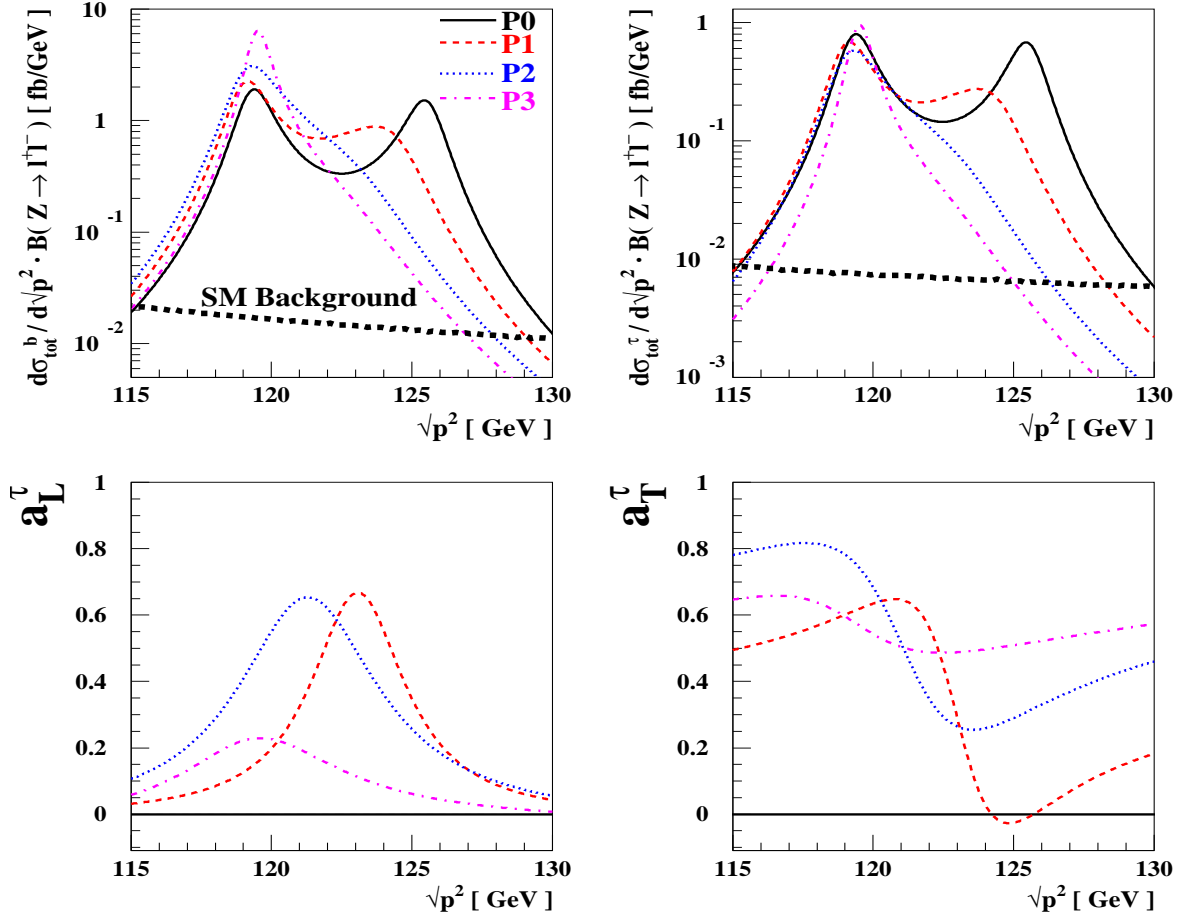,height=14cm,width=16cm}}
\vspace{-0.5cm}
\caption{{\small \it The   differential  total  cross section  
$d\hat{\sigma}^f_{\rm tot}(e^-e^+\to Z H_i \to Z \bar{f}f)/d\sqrt{p^2}$
multiplied by $B(Z\rightarrow l^+l^-)=B(Z\rightarrow e^+e^-)+B(Z\rightarrow \mu^+\mu^-)$  
as functions of the invariant mass of the Higgs decay products
$\sqrt{p^2}$  in units of fb/GeV when $f=b$ (upper-left panel) and $f=\tau$ (upper-right panel). 
The CP-conserving     
two-way mixing ({\bf P0})    and three
CP-violating tri-mixing ({\bf P1}-{\bf P3}) scenarios have been taken.
The lower two frames show the CP-violating asymmetries when Higgs
bosons decay into tau leptons.  The asymmetries are defined using the
longitudinal (lower-left panel) and transverse (lower-right panel)
polarizations of $\tau$ leptons.
See \cite {Ellis:2005ik} for details.
}}
\vspace{-0.4cm}
\label{fig:ilc}
\end{figure}

\noindent
The cross section of each process is given by
\begin{eqnarray}
\sigma(e^+e^-\rightarrow Z H_i)&=&g_{H_iVV}^2\,
\sigma_{\rm SM}^{HZ}(M_{H_{\rm SM}}\rightarrow M_{H_i})\,, \nonumber \\
\sigma(e^+e^-\rightarrow \nu\nu H_i)&=&g_{H_iVV}^2\,
\sigma^{WW}_{\rm SM}(M_{H_{\rm SM}}\rightarrow M_{H_i})\,, \nonumber \\
\sigma(e^+e^-\rightarrow H_i H_j)&=&g_{H_iH_jV}^2\,
\frac{G_F^2 M_Z^4}{6\pi s}(v_e^2+a_e^2)
\frac{\lambda^{3/2}(1,M_{H_i}^2/s,M_{H_j}^2/s)}{(1-M_Z^2/s)^2}\,,
\end{eqnarray}
where $g_{H_iH_jV}={\rm sign}[{\rm det}(O)]\epsilon_{ijk}\,g_{H_kVV}$,
$v_e=-1/4+s_W^2\,,a_e=1/4$ and $\sigma^{HZ\,,WW}_{\rm SM}$ denotes the corresponding
production cross section of the SM Higgs boson. As is well known,
the $WW$ fusion cross section grows as
$\ln(s)$ compared to the Higgsstrahlung and becomes dominant for large
center-of-mass energy $\sqrt{s}$. In the decoupling limit, $M_{H^\pm} \gsim 200$ GeV,
the couplings for heavier Higgs bosons $g_{H_{2,3}VV}$ are suppressed.
In this case, for the production of $H_2$ and $H_3$, the
pair production mechanism is active since $|g_{H_2H_3V}|=|g_{H_1VV}| \sim 1$.
When $M_{H^\pm} \lsim 200$ GeV,
the excellent energy and momentum
resolution of electrons and muons coming from measurements of
the $Z$ boson in Higgsstrahlung may help
to resolve a coupled system of neutral Higgs bosons by
analyzing the production lineshape, see the two upper panels of Fig.~\ref{fig:ilc}.

In the CP-invariant MSSM framework, there is a selection rule that only two CP-even 
Higgs bosons can be produced {\it via} the Higgsstrahlung process and only two 
combinations of 
CP-even and CP-odd Higgs bosons can be produced.
In other words, the observation of three distinct Higgs bosons
in Higgsstrahlung and $WW$ fusion and/or of all three pairs of Higgs bosons
in pair production could be interpreted as
a signal of CP violation in the MSSM framework. 
However, such an interpretation relies on the hypothesis that there
exist no additional singlet or doublet Higgs fields.
To confirm the existence of genuine CP violation, one needs to
measure other observables such as CP asymmetries.  In this light, the
final fermion spin-spin correlations in Higgs decays into tau leptons,
neutralinos, charginos, and top quarks need to be investigated.  
For the Higgsstrahlung process 
in which the produced Higgs bosons decay into a fermion pair $f\bar{f}$,
the differential cross-section for the Higgsstrahlung process  is given by
\begin{eqnarray}
p^2\frac{d\sigma}{dp^2}&=&
{\sigma^{HZ}_{\rm SM}(p^2)}\times
\frac{N_f\,g_f^2 \beta_f(p^2)}{16\pi^2}
\left\{
(1+{P_L\bar{P}_L})C^f_1(p^2)+
{(P_L+\bar{P}_L)}{C^f_2(p^2)} \right.\nonumber \\
&&+\left.
{P_T\bar{P}_T}\left[\cos(\alpha-\bar\alpha)C^f_3(p^2)
+\sin(\alpha-\bar\alpha){C^f_4(p^2)}\right] \right\}\,,
\label{eq:cx}
\end{eqnarray}
where $p^2$ and $N_f$ are the invariant mass squared 
and the color factor of the final-state fermions, respectively.
\begin{figure}[t]
\vspace*{1cm}
\begin{center}
\begin{picture}(300,100)(0,0)

\Line(15,10)(55,90)
\Line(55,90)(285,90)
\Line(285,90)(245,10)
\Line(245,10)(15,10)

\Text(-3,50)[]{$\tau^+$}
\Line(15,50)(145,50)
\Line(15,50)(20,54)
\Line(15,50)(20,47)

\Text(298,50)[]{$\tau^-$}
\Line(155,50)(285,50)
\Line(285,50)(280,54)
\Line(285,50)(280,47)

\Vertex(150,50){3}

\DashLine(90,80)(95,110){3}
\DashLine(85,50)(115,110){3}
\ArrowArc(85,50)(50,63,82)
\Text(102,105)[]{$\bar{\alpha}$}

\DashLine(225,50)(245,90){3}
\ArrowArc(225,50)(20,60,90)
\Text(233,77)[]{$\alpha$}

\SetWidth{2}
\Line(225,50)(225,80)\ArrowLine(225,80)(225,81)
\Text(225,99)[]{$P_{_T}$}

\Line(225,50)(255,50)\ArrowLine(255,50)(256,50)
\Text(240,40)[]{${P}_{_L}$}

\Line(85,50)(90,80)\ArrowLine(90,80)(91,86)
\Text(80,75)[]{$\bar{P}_{_T}$}

\Line(85,50)(50,50)\ArrowLine(50,50)(49,50)
\Text(65,40)[]{$\bar{P}_{_L}$}

\SetWidth{0.5}
\Line(225,50)(265,100)\ArrowLine(265,100)(266,101)\Text(275,110)[]{$h^-$}
\ArrowArc(225,50)(35,0,50)\Text(265,70)[]{$\theta^-$}
\DashLine(150,50)(265,100){3}
\Line(225,50)(185,0)\ArrowLine(185,0)(184,-1)\Text(180,-10)[]{$\nu_\tau$}
\DashLine(150,50)(185,0){3}

\DashLine(150,50)(45,115){3}
\DashLine(150,50)(115,0){3}
\Line(45,115)(115,0)\ArrowLine(45,115)(44,113)\ArrowLine(115,0)(116,-2)
\Text(40,105)[]{$\bar{\nu}_\tau$}
\Text(120,-10)[]{$h^+$}
\ArrowArcn(85,50)(20,0,300)\Text(114,40)[]{$\theta^+$}

\end{picture}\\
\end{center}
\medskip
\noindent
\caption{\it The $\tau^+\tau^-$ production plane in the Higgs-boson
rest frame, in the case when the $\tau$ leptons decay into hadrons
$h^\pm$ and neutrinos.  The longitudinal-polarization vector
$P_L(\bar{P}_L)$ and the transverse-polarization vector
$P_T(\bar{P}_T)$ with the azimuthal angle $\alpha(\bar\alpha)$ of
$\tau^-\,(\tau^+)$ are shown. }
\label{fig:tautau}
\end{figure}
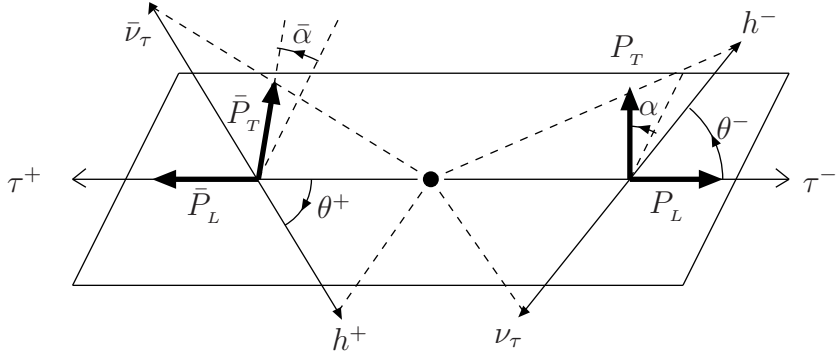
%
The polarization coefficients $C_i^f$ may conveniently be expressed as
follows:
\begin{eqnarray}
C^f_1(p^2) \!&=&\! \frac{1}{4}\left(\left|\langle + \rangle^f\right|^2
+\left|\langle - \rangle^f\right|^2\right)\,, \qquad\quad
C^f_2(p^2)\ =\ \frac{1}{4}\left(\left|\langle + \rangle^f\right|^2
-\left|\langle - \rangle^f\right|^2\right)\,, \nonumber \\
C^f_3(p^2) \!&=&\! -\frac{1}{2}\real\left(\langle + \rangle^f
\langle - \rangle^{f*}\right)\; ,  \qquad\qquad
C^f_4(p^2)\ =\ \frac{1}{2}\imag\left(\langle + \rangle^f
\langle - \rangle^{f*}\right)\; ,
\end{eqnarray}
with
\begin{equation}
\langle\sigma\rangle^f = \sum_{i,j}
g_{H_iVV}D_{ij}(p^2)\, (\sigma\beta
g^S_{H_j\bar{f}f}-ig^P_{H_j\bar{f}f})\,.
\end{equation}
When $f=\tau$ and tau leptons subsequently
decay into  charged hadrons $h^\mp$ and neutrinos, 
see Fig.~\ref{fig:tautau}, we have
\begin{equation}
P_L=\cos\theta^-\,, \
P_T=\sin\theta^-\,, \
\alpha=\varphi^-\,; \ \
\bar{P}_L=\cos\theta^+\,, \
\bar{P}_T=\sin\theta^+\,, \
\bar{\alpha}=\varphi^+-\pi\,,
\end{equation}
by identifying the polarization analyser for $\tau^\mp$ as
$\hat{a}^\mp=\pm\hat{h}^{\mp}$ with $\hat{h}^\mp$ denoting unit vectors parallel 
to the $h^\mp$ momenta in the $\tau^\mp$ rest frame. 
The angles  $\theta^\pm$ and   $\varphi^\pm$ are  the polar  and  azimuthal
angles of  $h^\pm$, respectively, in the  $\tau^\pm$ rest frame.  This implies that
the  polarization coefficients $C_i^\tau(p^2)$ can be
determined  by  examining the   angular distributions of   the charged
hadrons coming  from  the $\tau$-lepton  decays.
See the lower panels of Fig.~\ref{fig:ilc} for the CP asymmetries
$a^\tau_{L}=C^\tau_{2}/C^\tau_1$ and
$a^\tau_{T}=C^\tau_{4}/C^\tau_1$ \cite{Ellis:2005ik}.
%

\subsection{$\gamma$LC}
The two-photon collider option of the ILC, the $\gamma$LC, offers
unique capabilities for probing  CP violation in the MSSM
Higgs sector, because one  may vary the initial-state polarizations as
well as measure  the   polarizations of  some  final  states  in Higgs
decays  \cite{Ellis:2004hw}. The amplitude contributing to 
$\gamma(\lambda_1)\gamma(\lambda_2)\rightarrow H \rightarrow f(\sigma)\bar{f}(\bar{\sigma})$
is given by
\begin{equation}
{\cal M}_H \; = \; \frac{\alpha \, m_f \, \sqrt{\hat{s}}}{4\pi v^2}
\langle \sigma ;\lambda_1 \rangle^f_H
\delta_{\sigma\bar{\sigma}}\delta_{\lambda_1\lambda_2},
\label{eq:MH}
\end{equation}
where the reduced amplitude
\begin{equation}
\langle \sigma ;\lambda \rangle^f_H \; = \; \sum_{i,j=1}^3
[S_i^\gamma(\sqrt{\hat{s}})+i\lambda P_i^\gamma(\sqrt{\hat{s}})]
\ D_{ij}(\hat{s}) \
(\sigma\beta_f g^S_{H_j\bar{f}f}-ig^P_{H_j\bar{f}f}),
\label{eq:redHamp}
\end{equation}
is a quantity given by the Higgs-boson propagator matrix
Eq.~(\ref{eq:Hprop}) combined with the production and decay vertices.
The one-loop induced complex couplings of the $\gamma \gamma H_i$
vertex, $S_i^\gamma(\sqrt{\hat{s}})$ and $P_i^\gamma(\sqrt{\hat{s}})$,
get dominant contributions from charged particles such as the bottom
and top quarks, tau leptons, $W^\pm$ bosons, charginos,
third-generation sfermions and charged Higgs bosons
\footnote{The arrays
{\tt SPHO(3)} and {\tt PPHO(3)} are used for
the $\hat{s}$-dependent $\gamma$-$\gamma$-Higgs couplings.} :
\begin{eqnarray}
S^\gamma_i(\sqrt{\hat{s}})
 &=& 2\!\!\sum_{f=b,t,\tilde{\chi}^\pm_{1,2}} N_C\, Q_f^2\,
     g_fg^{S}_{H_i\bar{f}f}\,\frac{v}{m_f} F_{sf}(\tau_{f})
    \, -\!\!\!\!\!\!\!\! \sum_{\tilde{f}_j=\tilde{t}_{1,2},\tilde{b}_{1,2},
            \tilde{\tau}_{1,2}}\!\!\!\!\!\!\!\!\! N_C\, Q_f^2g_{H_i\tilde{f}^*_j
            \tilde{f}_j} \frac{v^2}{2m_{\tilde{f}_j}^2}
            F_0(\tau_{\tilde{f}_j}) \nonumber \\
  && -g_{_{H_iVV}}F_1(\tau_{W})- g_{_{H_iH^+H^-}}\frac{v^2}{2 M_{H^\pm}^2}
      F_0(\tau_{H^\pm}) \,, \nonumber \\
P^\gamma_i(\sqrt{\hat{s}})
  &=& 2\!\!\sum_{f=b,t,\tilde{\chi}^\pm_{1,2}} N_C\,Q_f^2\,g_f
      g^{P}_{H_i\bar{f}f} \,\frac{v}{m_f} F_{pf}(\tau_{f})\,,
\end{eqnarray}
where $\tau_{x}=\hat{s}/4m_x^2$, $N_C=3$ for (s)quarks and $N_C=1$ for
staus and charginos, respectively.  
For the explicit forms of $F_{1}$ and couplings, see \cite{Lee:2003nt}.

One advantage of the $\gamma$LC option over $e^+e^-$ collisions at the ILC is
that one can construct CP asymmetries even when Higgs bosons decay
into muons and $b$ quarks, by exploiting the controllable beam
polarizations of the colliding photons. 
\begin{figure}[t]
\vspace{-1.3cm}
\centerline{\epsfig{figure=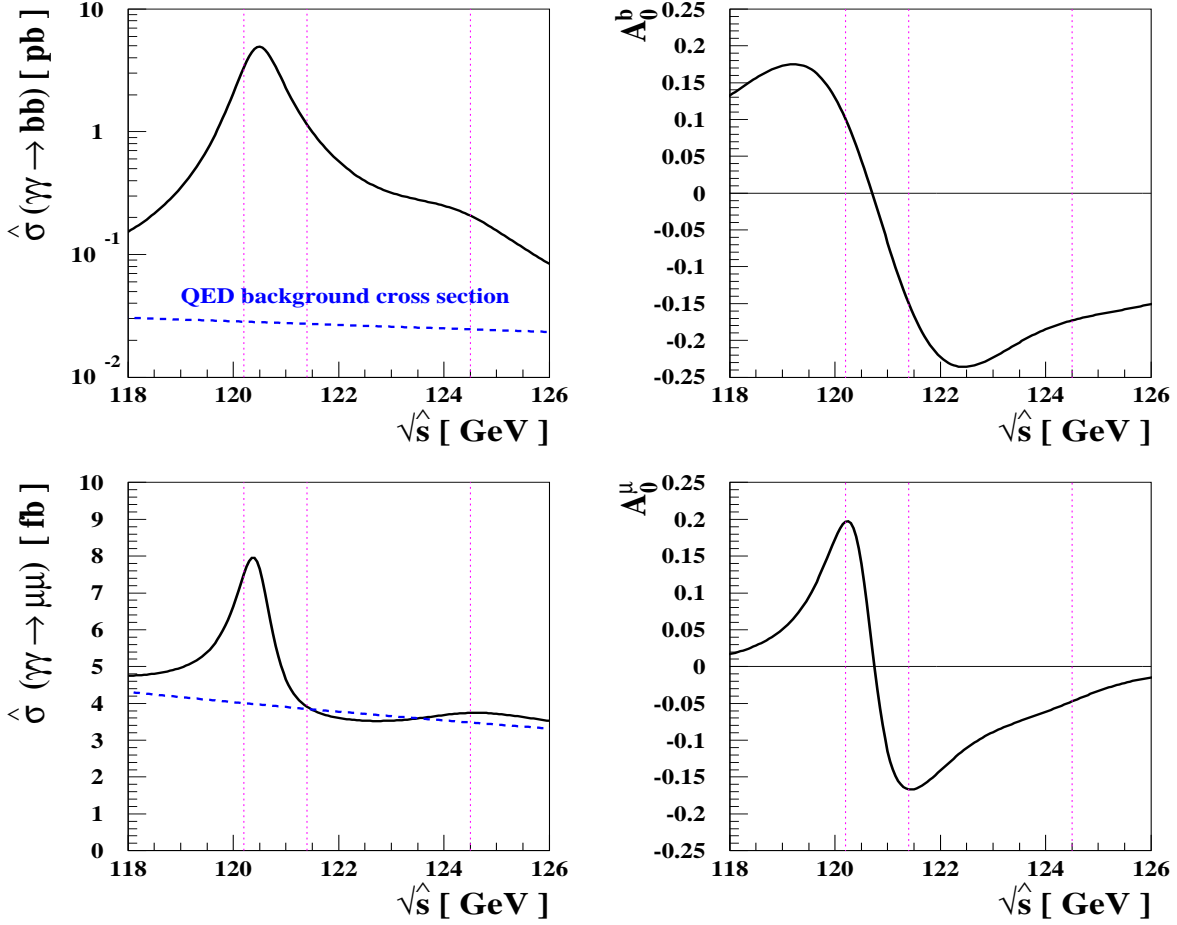,height=14cm,width=16cm}}
\vspace{-0.5cm}
\caption{\small \it The cross sections (left column) and the CP asymmetries ${\cal
A}^f_0$ (right
column) for the processes $\gamma\gamma \to \bar{b}b$ (upper) and
$\gamma\gamma \to \mu^+\mu^-$ (lower).
The QED continuum contributions to the cross sections are also shown.
The tri-mixing scenario with $\Phi_3=-10^\circ$ and $\Phi_A=90^\circ$ has been
considered, making the angle cuts $\theta^b_{\rm cut}=280$ mrad and
$\theta^\mu_{\rm cut}=130$ mrad. The three Higgs
masses are indicated by vertical lines.
See \cite {Ellis:2004hw} for details.  }
\label{fig:plc}
\end{figure}
In the case of identical photon 
helicities $\lambda_1=\lambda_2=\lambda$, 
corresponding to the different combinations of helicities of the
initial-state photons $\lambda$ and final-state fermions $\sigma$, 
we have four cross sections:
\begin{equation}
\hat{\sigma}^f_{\sigma\lambda}(\gamma\gamma\rightarrow f \bar{f})
=\frac{\beta_f N_C}{32 \pi}
\left(\frac{\alpha \, m_f }{4\pi v^2} \right)^2
{\cal Y}^f_{\sigma\lambda}
\end{equation}
where $\sigma\,,\lambda=\pm\,$ and
\begin{eqnarray}
{\cal Y}^f_{\sigma\lambda}\equiv
2\left|\langle \sigma;\lambda \rangle^f_H \right|^2
+2R(\hat{s})^2\, F_1^{z_f} \left|\langle \sigma;\lambda \rangle^f_C \right|^2
+2R(\hat{s})\, F_2^{z_f} \ \real(\langle \sigma;\lambda \rangle^f_H
\langle \sigma;\lambda \rangle_C^{f*}) \,.
\label{eq:y}
\end{eqnarray}
We have included the QED continuum contribution
and an experimental cut on the fermion polar angle
$\theta$ has been introduced: 
$|\cos\theta| \leq z_f$ and $\cos\theta^f_{\rm cut}=z_f$.
For the explicit forms of the functions $F_1^{z_f}$, $F_2^{z_f}$, $R(\hat{s})$,
and the QED amplitude $\langle \sigma;\lambda \rangle^f_C$, 
we refer to Ref.~\cite{Ellis:2004hw}. In this case, 
one can construct CP asymmetries with no need to determine the helicities
of the final states as
\begin{equation}
{\cal A}^f_0\equiv
\frac{\sum_{\sigma=\pm}(\hat{\sigma}^f_{\sigma +}-\hat{\sigma}^f_{\sigma -})}
{4\,\hat{\sigma}(\gamma\gamma\to f\bar{f})}\,,
\end{equation}
where
$\hat{\sigma}(\gamma\gamma\to f\bar{f})=\left(
\hat{\sigma}^f_{++}+\hat{\sigma}^f_{--}+
\hat{\sigma}^f_{+-}+\hat{\sigma}^f_{-+} \right)/4$.
Figure \ref{fig:plc} shows the cross section $\hat{\sigma}(\gamma\gamma\to f\bar{f})$ 
and the CP asymmetry ${\cal A}^f_0$ when $f=b$ and $\tau$.

By considering the interference of Higgs-mediated resonant and QED
continuum amplitudes and the possibly measurable final-state
polarizations, one can construct more than 20 independent observables
for each decay mode and half of them are CP odd.
This suggests that one can investigate all possible spin-spin
correlations in the final states such as tau leptons, neutralinos,
charginos, top quarks, vector bosons, etc., with the goal of complete
determination of CP-violating Higgs-boson couplings to them. The cases
of tau-lepton and top-quark final states are demonstrated in \cite
{Ellis:2004hw}.

\subsection{MC}

For the complete determination of the CP-violating Higgs-boson couplings
to SM as well as supersymmetric particles, a MC is even
better than the $\gamma$LC.
At a muon collider, it is possible to
control the energy resolution and polarizations of both the muon and
the anti-muon.  Compared to the $\gamma$LC case, the center-of-mass
frame is known very accurately, providing much better resolving power for a
nearly-degenerate system of Higgs bosons.

For example, let us consider the process 
$\mu^-(\lambda)\mu^+(\bar{\lambda})
\rightarrow f(\sigma)\bar{f}(\bar{\sigma})$.
When the helicities of the muons are equal: $\lambda=\bar{\lambda}$,
the $\gamma$- and $Z$-mediated processes 
are suppressed by a kinematical factor of the muon mass, 
and the Higgs-mediated amplitude contributing to the process
is given by 
\begin{equation}
{\cal M}_H \; = \; -g_\mu\,g_f
\langle \sigma ;\lambda \rangle^f_\mu\,
\delta_{\sigma\bar{\sigma}}\delta_{\lambda\bar{\lambda}}\,,
\end{equation}
where $g_f=gm_f/2M_W$. The reduced amplitude is
\begin{equation}
\langle \sigma ;\lambda \rangle^f_\mu \; = \; \sum_{i,j=1}^3
(\lambda\beta_\mu\, g^S_{H_i\bar{\mu}\mu}+ig^P_{H_i\bar{\mu}\mu})
\ D_{ij}({s}) \
(\sigma\beta_f g^S_{H_j\bar{f}f}-ig^P_{H_j\bar{f}f})\,,
\end{equation}
where $s$ is the invariant muon-cillider energy squared.
The observables obtained by controlling solely the muon and anti-muon
polarizations are given by \cite{ACL_mumu}
\begin{equation}
\sigma_{_{RR/LL}}  = \frac{g_\mu^2\,g_f^2\,N_C\, \beta_f}{8\,\pi\,s}\,
              \,\left[\, C^f_3\pm C^f_4\,\right]\,,  \ \ \ \
\sigma_{\,\Vert/\bot}= \frac{g_\mu^2\,g_f^2\,N_C\, \beta_f}{16\,\pi\,s}\,
(\pm)\,C^f_{15/16}\,,
\end{equation}
where the coefficients are
\begin{eqnarray}
C^f_3&=&\frac{1}{4}\sum_{\sigma=\pm}
\left[|\langle\sigma;+\rangle^f_\mu|^2+|\langle\sigma;-\rangle^f_\mu|^2\right]\,,
\ \ \ \
C^f_4=\frac{1}{4}\sum_{\sigma=\pm}
\left[|\langle\sigma;+\rangle^f_\mu|^2-|\langle\sigma;-\rangle^f_\mu|^2\right]\,,
\nonumber \\
C^f_{15}&=&-\frac{1}{2}\,\real\,\sum_{\sigma=\pm}
\left[\langle\sigma;-\rangle^f_\mu\langle\sigma;+\rangle^{f\,*}_\mu\right]\,,
\ \ \ \ \
C^f_{16}=\frac{1}{2}\,\imag\,\sum_{\sigma=\pm}
\left[\langle\sigma;-\rangle^f_\mu\langle\sigma;+\rangle^{f\,*}_\mu\right]\,.
\end{eqnarray}
We note that $C^f_4$ and $C_{16}^f$ are CP-odd observables.
\begin{figure}[t]
\vspace{-2.0cm}
\centerline{\epsfig{figure=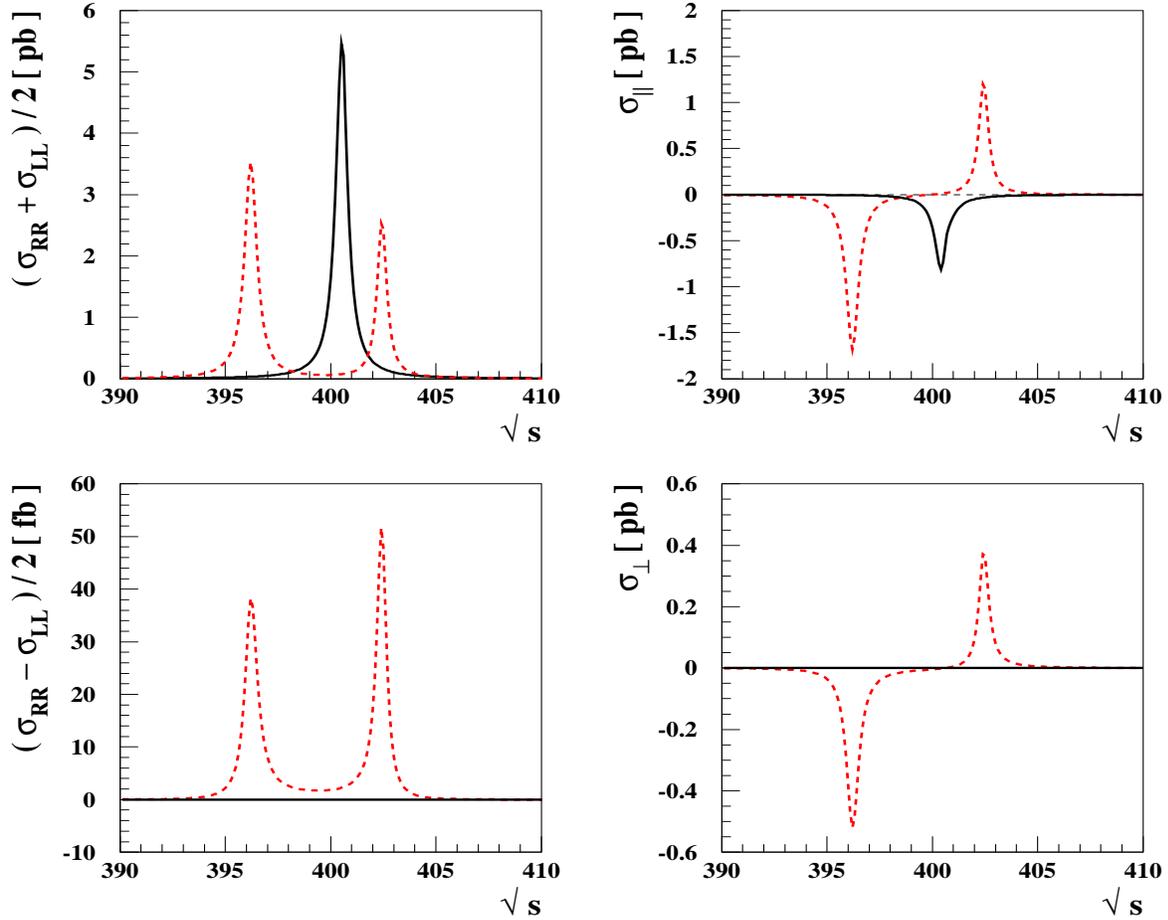,height=14cm,width=16cm}}
\vspace{-0.5cm}
\caption{\small \it The cross sections  
$(\sigma_{_{RR}}+\sigma_{_{LL}})/2$ (upper left), 
$\sigma_{\,\Vert}$ (upper right),
$(\sigma_{_{RR}}-\sigma_{_{LL}})/2$ (lower left),
and $\sigma_{\bot}$ (lower right) for the process
$\mu^-\mu^+ \rightarrow H \rightarrow t\bar{t}$
as functions of $\sqrt{s}$. Here, $(\sigma_{_{RR}}\pm\sigma_{_{LL}})/2$ is 
given in units of fb but the other cross sections are in pb.
The solid and dashed lines are for $\Phi_A=0^\circ$
and $\Phi_A=-90^\circ$, respectively.
}
\vspace{-0.3cm}
\label{fig:mmtt}
\end{figure}
In Fig.~\ref{fig:mmtt}, we show $(\sigma_{_{RR}}\pm\sigma_{_{LL}})/2$  and
$\sigma_{\,\Vert/\bot}$ when Higgs bosons decay into top quarks $f=t$, 
as functions of $\sqrt{s}$. The parameters are chosen
as in Ref.~\cite{ACL_mumu}, except that $\tan\beta=5$ and $M_{H^\pm}=407$ GeV.
The solid lines are for the CP-conserving
case with $\Phi_A=0^\circ$ and the dashed lines are for
the CP-violating case with $\Phi_A=-90^\circ$.
The masses are 
$M_{H_2}=400.5$ GeV and $M_{H_3}=400.6$ GeV in the CP-conserving case,
but we have a larger mass splitting in the CP-violating case: $M_{H_2}=396.2$ GeV 
and $M_{H_3}=402.4$ GeV. 
The results have been updated,
compared to those presented in Ref.~\cite{ACL_mumu}, 
by including the off-diagonal absorptive parts in the
Higgs-boson propagators, Higgs-boson pole mass shifts, 
$b$-quark Yukawa-coupling resummation effects, etc.
More observables could be obtained by considering final-state spin correlations and
interference effects with the processes mediated by $\gamma$ and $Z$
bosons.

\section{Low-energy Observables}

Low-energy observables provide indirect constraints on the soft SUSY
breaking parameters. The observables are particularly useful for
identifying the favoured range of parameter space when the SM
predictions for them are strongly suppressed and/or precise
experimental measurements of them have been performed.  Such
observables include EDMs, $(g-2)_\mu$, 
${\rm BR}(b\rightarrow s\gamma)$,
${\cal A}_{\rm CP}(b\rightarrow s\gamma)$,
${\rm BR}(B \rightarrow K l\,l)$ and ${\rm BR}(B_{s,d} \rightarrow l^+ l^-)$.

\begin{figure}[t]
\vspace{-2.0cm}
\centerline{\epsfig{figure=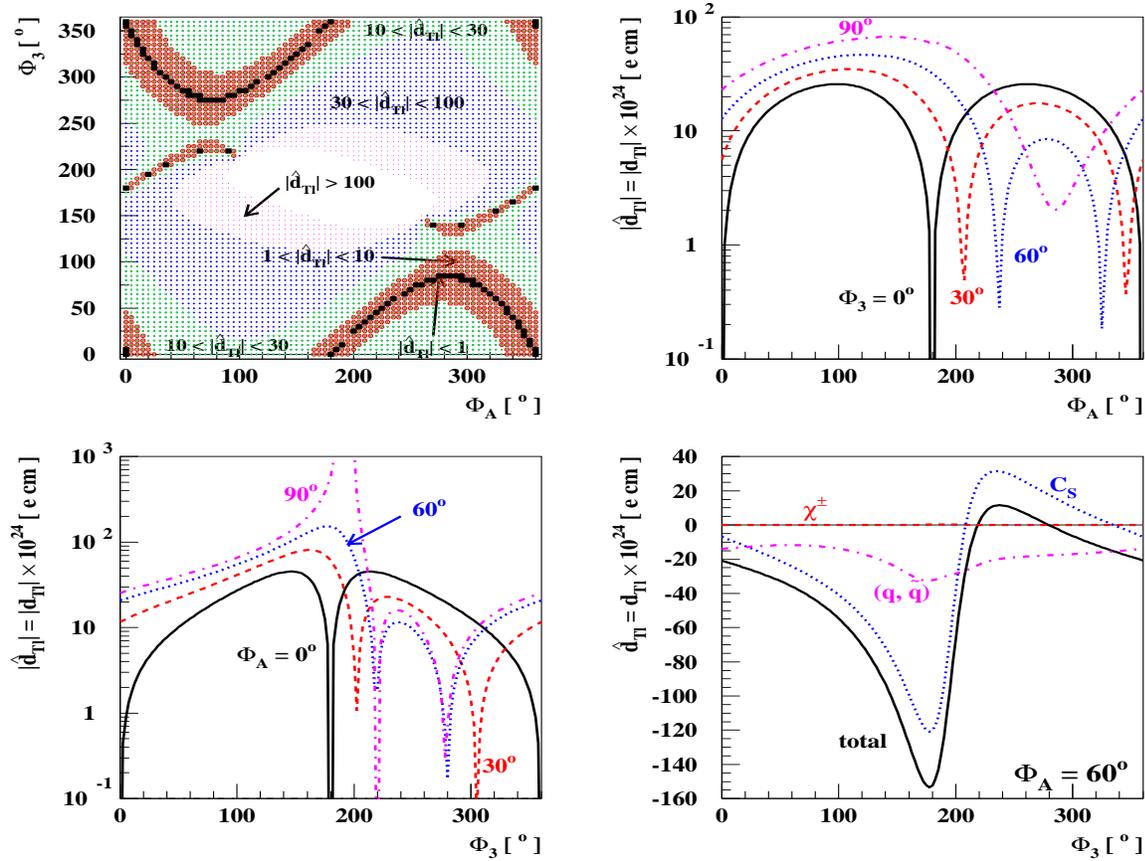,height=13cm,width=16cm}}
\vspace{-0.5cm}
\caption{\small \it The Thallium EDM $\hat{d}_{\rm Tl} \equiv d_{\rm Tl} \times
10^{24}$~$e\,cm$ in the tri-mixing scenario. The
upper-left frame displays $|\hat{d}_{\rm Tl}|$ in the $(\Phi_A,
\Phi_3)$
plane. The unshaded region around the point $\Phi_3=\Phi_A=180^o$ is
not theoretically allowed.
The different shaded regions correspond to different ranges of
$|\hat{d}_{\rm Tl}|$, as shown: specifically, $|\hat{d}_{\rm Tl}|<1$
in the narrow region denoted by filled black squares.
In the upper-right frame, we show $|\hat{d}_{\rm Tl}|$ as a function of
$\Phi_A$ for several values of $\Phi_3$.
In the lower-left frame, we show $|\hat{d}_{\rm Tl}|$ as a function of
$\Phi_3$ for four values of $\Phi_A$.
In the lower-right frame, we show the $C_S$ (dotted line) and
$d_e$ (dash-dotted line) contributions to
$\hat{d}_{\rm Tl}$ separately
as functions of $\Phi_3$ when $\Phi_A=60^o$. As shown by the dashed
line, the chargino contribution is negligible.
See \cite {Ellis:2005ik} for details.  }
\vspace{-0.3cm}
\label{fig:th.edm}
\end{figure}

Currently, the EDM of the Thallium atom provides one of the best
constraints on the CP-violating phases, depending on the SUSY
scale. The main contributions to the atomic EDM of $^{235}$Tl come
from two terms. One of them is the electron EDM $d_e$ and the other
is the coefficient $C_S$ of a CP-odd electron-nucleon interaction. The
coefficient $C_S$ is essentially given by the gluon-gluon-Higgs
couplings and the two-loop Higgs-mediated electron
EDM~\cite{Chang:1998uc,Pilaftsis:2002fe}  
is given by the
sum of contributions from third-generation quarks and squarks and
charginos
\footnote{Here we have not considered one-loop contributions, since
they are independent in the absence of any assumptions relating the
soft SUSY-breaking terms of the first two generations to those of the
third generation.}.  We show in \cite{Ellis:2005ik} that it is
straightforward to obtain $C_S$ and the Higgs-mediated $d_e$ by use of
couplings calculated by {\tt CPsuperH}. 
As can be seen for example in Fig.~\ref{fig:th.edm},
one can then implement the
thallium EDM constraint on the CP-violating phases and demonstrate
that they leave open the possibility of large CP-violating effects in
Higgs production at the ILC. 

The code  {\tt CPsuperH}  will be extended  in near future  to include
CP-violating  effective  FCNC  Higgs-boson  interactions  to  up-  and
down-type
quarks~\cite{Dedes:2002er,Feng:2006ze,Gomez:2006uv,Carena:2006ai}.
The  determination  of these  effective  interactions  may be  further
improved in  the framework of  an effective potential  approach, where
the most significant subleading  contributions to the couplings can be
consistently   incorporated.    At   large  values   of   $\tan\beta$,
Higgs-mediated interactions contribute  significantly to the $B$-meson
observables mentioned above and so  may offer novel constraints on the
parameter  space of  constrained versions  of  the MSSM,  such as  the
scenario of minimal flavour violation.

\section{Cosmological Connections and Future Directions}

Imposing $R$ parity, the Lightest Supersymmetric Particle (LSP) is
stable and becomes one of the strongest candidates for the cold dark
matter which provides some 23 \% of the energy density of the
Universe. When the lightest neutralino is the LSP, the annihilation of
neutralino pairs and scattering off ordinary matter become relevant in
Cosmology.
Neutralino-pair annihilation cross sections are needed in calculations
of the cosmological neutralino relic density and of the fluxes of
cosmic-ray antiprotons, positrons and gamma-rays due to relic
annihilations in the Galactic halo. The fluxes of high-energy
neutrinos from the cores of the Sun and Earth depend on the scattering
processes leading to neutralino capture, as well as on the rates
for different annihilation channels. Scattering cross sections are
also important for the direct detection of the neutralino via its
interaction with matter.

In the presence of CP-violating mixing in Higgs sector, the couplings
of Higgs bosons to the SM and SUSY particles are significantly
modified and the contributions to the pair-annihilation and scattering
cross sections from Higgs mediated processes could be largely
affected. For example, for non-relativistic neutralinos, the amplitude
of the process $\tilde{\chi}^0_1\tilde{\chi}^0_1\rightarrow H_i
\rightarrow WW\,, H_1H_1$ is highly suppressed in CP-invariant
theories, but may be enhanced when the intermediate Higgs bosons are
CP-mixed states \cite{Nihei:2005va}.
We note that CPsuperH has been used for the calculation of the
relic abundance of the lightest neutralino in Ref.~\cite{Belanger:2006qa}.

As mentioned above, CP-violating phases in the MSSM may also make
possible baryogenesis at the electroweak
scale~\cite{Carena:2002ss,Konstandin:2005cd,Cirigliano:2006wh,Cirigliano:2006dg}. A key
ingredient in verifying this and other cosmological connections will
be laboratory measurements of CP-violating phenomena, and {\tt
CPsuperH} will be  a key tool for evaluating their implications for the
underlying model parameters by extending its scope to low-energy and cosmological
observables.

\section*{Acknowledgements}

The work of JSL was supported in part by The Korea Research
Foundation and The Korean Federation of Science and Technology Societies 
Grant funded by Korea Government (MOEHRD, Basic Research Promotion Fund).
The work of AP is supported in part by the PPARC research grant: PP/D000157/1.


\end{document}